\documentclass{iopart}
\usepackage{color}
\usepackage{amssymb}
\usepackage[dvips]{graphicx}

\begin{document}

\title[Determination of the angular momentum distribution of supernovae]{Determination of the angular momentum distribution of supernovae from gravitational wave observations}

\author{K~Hayama$^1$, S~Desai$^2$, K Kotake$^3$, S~D~Mohanty$^1$, M~Rakhmanov$^1$, T~Summerscales$^4$, S~Yoshida$^5$}
\address{$^1$The University of Texas at Brownsville, Brownsville, TX, 78520, US}
\address{$^2$ Center for Gravitational Wave Physics, Pennsylvania State University, University Park, PA 16802, US}
\address{$^3$Division of Theoretical Astronomy/Center for Computational Astrophysics, National Astronomical Observatory of Japan, 2-21-1, Osawa, Mitaka, Tokyo, 181-8588, Japan}
\address{$^4$Department of Physics, Andrews University, Berrien Springs, MI 49104, US}
\address{$^5$Southeastern Louisiana University, Hammond, LA 70402, US}

\ead{kazuhiro.hayama@ligo.org}

\begin{abstract} Significant progress has been made in the development of an international 
network of gravitational wave detectors, such as TAMA300, LIGO, VIRGO, and GEO600.
For these detectors, one of the most promising sources of gravitational waves are
core collapse supernovae especially in our Galaxy. 
Recent simulations of core collapse supernovae, rigorously 
carried out by various groups, show that the features of the waveforms are 
determined by the rotational profiles of the core, such as the rotation rate
 and the degree of the differential rotation prior to core-collapse. Specifically,
 it has been predicted that the sign of the second largest peak in the 
 gravitational wave strain signal is negative if the core rotates cylindrically with strong 
differential rotation. 
The sign of the second peak could be a nice indicator that provides us with 
information about the angular momentum distribution of the core, unseen without
gravitational wave signals.
Here we present a data analysis procedure aiming at the detection of the
second peak using a coherent network analysis and estimate the
detection efficiency when a supernova is at the sky location of the
Galactic center. 
The simulations showed we were able to determine the sign of the second peak under an idealized
condition of a network of gravitational wave detectors if a supernova occurs at the Galactic center.
\end{abstract}

\pacs{04.80.Nn, 07.05.Kf, 95.55.Ym, 95.85.Nv}

\section{Introduction}
Current estimates of Galactic core-collapse supernovae rates predict one 
core-collapse supernova event every $\sim 40 \pm 10$ year 
\cite{tammann:1994,Beacom:2001}. When a massive star 
($>\sim 10 M_{\odot}$) \cite{heger:2003} undergoes a core-collapse supernova explosion 
in our Galactic center, copious numbers of neutrinos are produced, some of which 
may be detected on the earth. 
Such supernova neutrinos will carry valuable information from deep inside the core. 
In fact, the detection of neutrinos from SN1987A (albeit in the Large Magellanic Cloud) 
paved the way for {\it Neutrino Astronomy}, an alternative to 
conventional astronomy by electromagnetic waves \cite{Hirata:87,SatoSuzuki:87,Beacom:1999}. 
Core-collapse supernovae are now poised to give birth to yet another
astronomy, {\it Gravitational-Wave Astronomy}.
 Currently long-baseline laser interferometers 
such as LIGO, VIRGO, GEO600, and TAMA300 are operational 
\cite{abbottetal:2004,Acerneseetal:2002,willke:2004,takahashi:2004}.
For these detectors, core-collapse supernovae have been proposed as one of 
the most plausible sources of gravitational waves (GWs) 
(see, e.g., \cite{new:2003,kotake:2006} for reviews).

If gravitational collapse of the supernova core proceeds 
spherically, no GWs can be emitted. 
To produce GWs, the gravitational collapse should proceed
 aspherically and dynamically. Observational evidence gathered over
the last few decades has pointed towards core-collapse supernovae indeed being
generally aspherical \cite{wang96,wang01,Wang:2002}. 
The most unequivocal example is SN1987A. Recent HST
images of SN1987A are directly showing that the expanding envelope is 
elliptical with the long axis aligned with the rotation axis inferred
from the ring. The aspect ratio and position angle of the symmetry axis are 
consistent with those predicted earlier from the observations of speckle and 
linear polarization. What is more, the linear polarization became greater as time 
passed \cite{wang01,leonard01}, a fact which has been used to argue that 
the central engine of the explosion is responsible for the
non-sphericity \cite{wheeler00,Maeda:2008}.
From a theoretical point of view, clarifying what makes the dynamics of the core
deviate from spherical symmetry is essential in understanding the GW emission
mechanism. 

It is well known that stars are generally rotating \cite{tass:1978} (see 
\cite{heger:05,heger:07} for recent controversy about the angular momentum transfer problem during stellar 
evolutions). 
This stellar rotation has been long supposed to play an
important role in the GW emission from core-collapse
supernovae, because the large-scale asphericities at core bounce 
induced by rotation can convert part of the gravitational energy 
into the form of GWs. Such GW signals 
are often simply called bounce signals (see \cite{muyan97,mueller,burohey,fryer04,kotake_gw_sasi} for other sources of
GWs from the postbounce phase such as convection and anisotropic neutrino 
radiations). 
Just recently, several investigations have been carried out that explore how 
astrophysical information can be obtained from supernovae GW waveforms
\cite{Hayama:2004,Summerscales:2008}. Hayama (2004) \cite{Hayama:2004} studied 
how accurately the waveforms associated with various supernova models may be estimated. 
Summerscales et al.(2008) \cite{Summerscales:2008} developed a maximum entropy approaches 
to extract the waveforms from noisy observations and discover what source information they contain.
 Such data analysis methods are not only 
indispensable for the first direct GW observation to confirm Einstein's theory of 
General Relativity, but also for clarifying supernova physics itself; 
such as the long-veiled explosion mechanism, hidden 
deep inside the core.

Theoretically, much attention has been paid to the bounce signals 
in the context of rotational (e.g., \cite{moenchgw,ZwergerMueller:1997,Kotake:2003,Ottetal:2004,
Dimmelmeier:2002B,dimmelprl} and references therein) and magneotrotational core-collapse 
\cite{kotakegwMHD,obergaulinger,CerdaDuranetal:2005} 
 with different levels of sophistication employed in the
 treatment of the equations of state (EOSs), neutrino transport,
 and general relativity (see \cite{kotake:2006} for a review). 
 Waveforms obtained in numerical simulations
 are often categorized to the three types, namely types I, II, and III. 
Type I waveforms are distinguished by a large amplitude peak at core bounce and
subsequent damping ring-down oscillations. Type I waveforms are obtained 
when the initial angular momentum is small, which leads to core bounce 
near nuclear densities. As the angular momentum of the core becomes larger with 
stronger differential rotation, the waveforms transit to type II 
waveforms, which have several distinct peaks caused by multiple
bounces supported by centrifugal forces. Type III, which are obtained for an 
extremely soft equation of state, are now considered to be unrealistic.

Through inspection of some of the elaborate numerical simulations cited above, 
it was suggested that the sign of the second 
largest peak becomes generically negative when the core 
rotates cylindrically with strong differential rotation \cite{Kotake:2003}. 
Looking carefully, this feature also can be seen in the type II 
waveforms obtained by other simulations 
(\cite{Ottetal:2004,ShibataSekiguchi:2004,CerdaDuranetal:2005}).
If we can observe the sign of the second peak, we could obtain information about the 
angular momentum distribution of an evolved massive star. The absolute amplitude of 
the second peak is within the sensitivity of initial LIGO for such a source at the 
galactic center. By using world-wide detector networks, the gravitational waveforms 
$h_+$, $h_\times$ can be reconstructed. In this paper, we look at how we might 
detect the second peak by 
using current LIGO and VIRGO. We propose a method to recover gravitational waveforms 
from supernova core collapse by using the {\tt RIDGE} coherent network analysis 
\cite{Hayama:2007}, which takes full advantage of multiple detectors. 
By carrying out Monte Carlo simulations, we estimate the detection efficiency 
of determining the correct sign of the second peak using an idealized LIGO-VIRGO network.

\section{Second peak of gravitational waves from core-collapse supernovae}
\begin{figure}
\begin{center}
\includegraphics[width=0.45\linewidth, height=4cm]{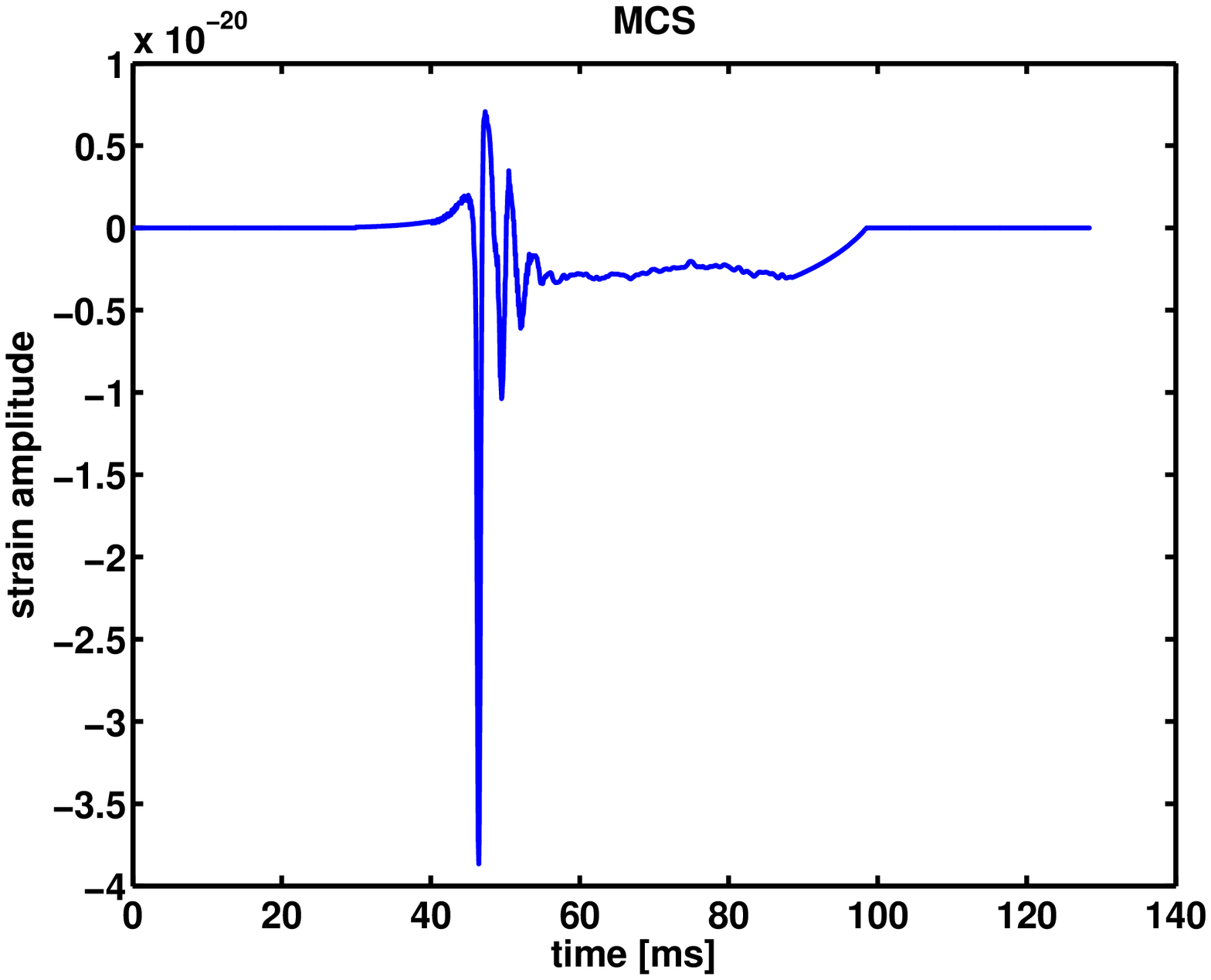}
\includegraphics[width=0.45\linewidth, height=4cm]{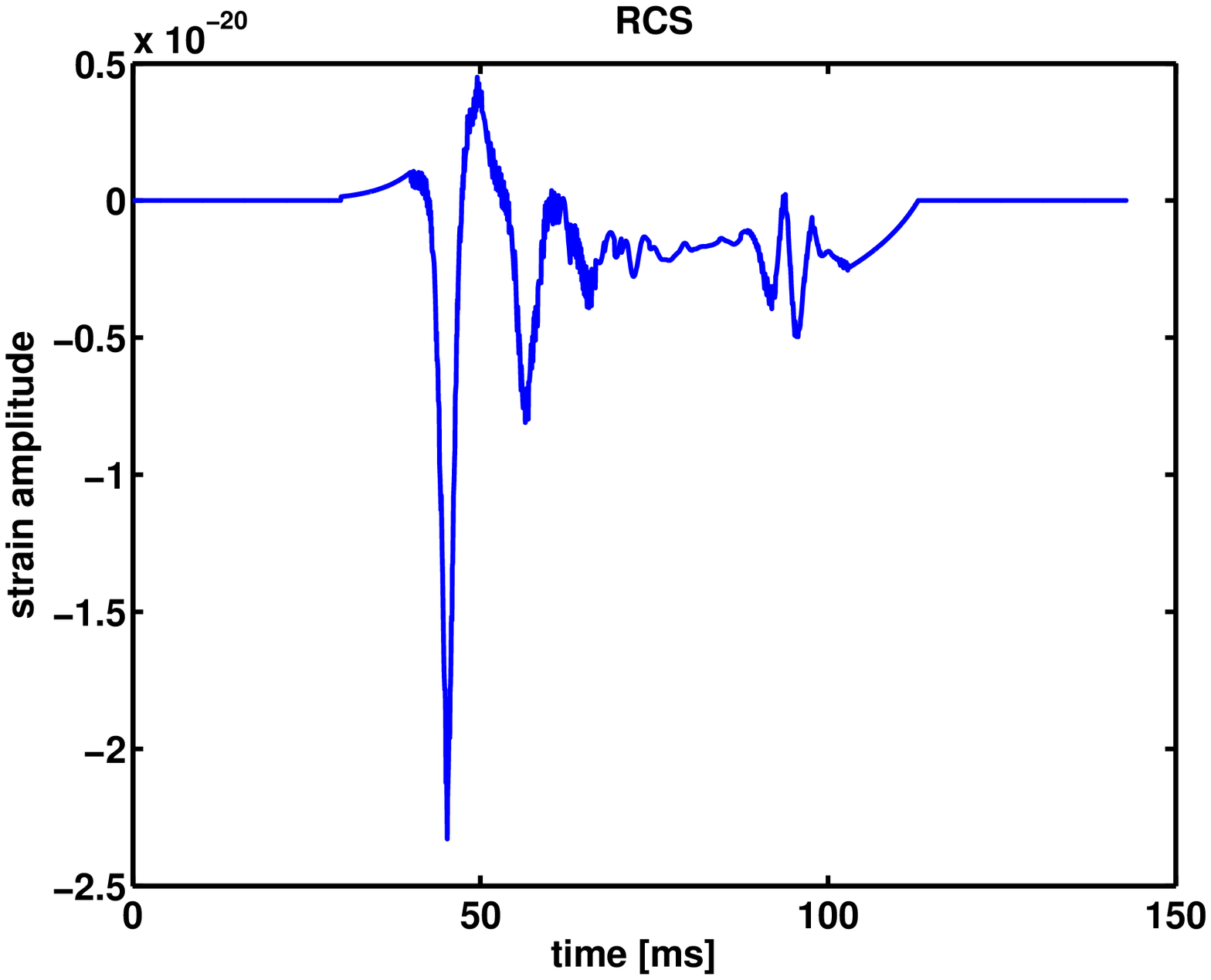}

\includegraphics[width=0.45\linewidth, height=4cm]{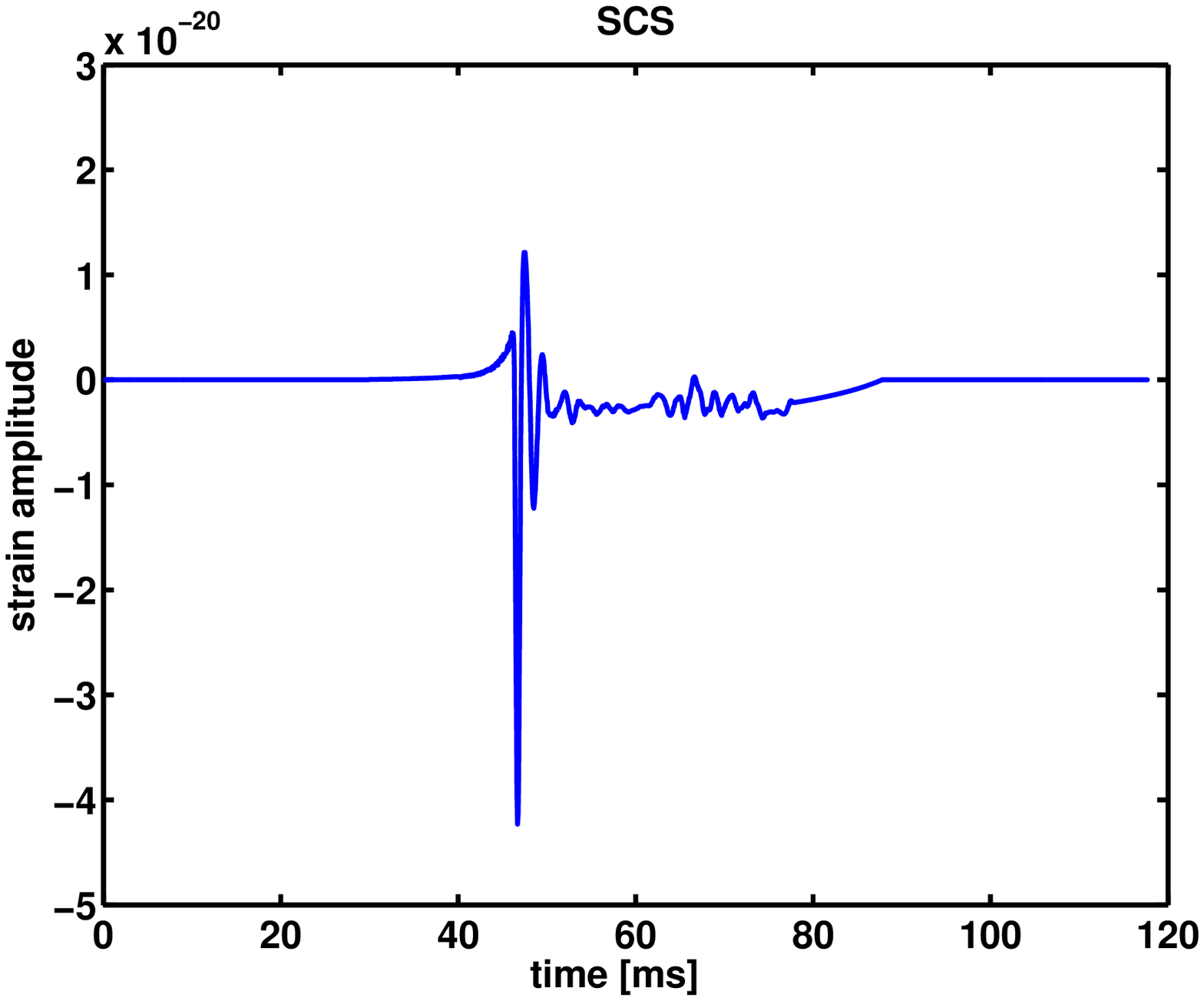}
\includegraphics[width=0.45\linewidth, height=4cm]{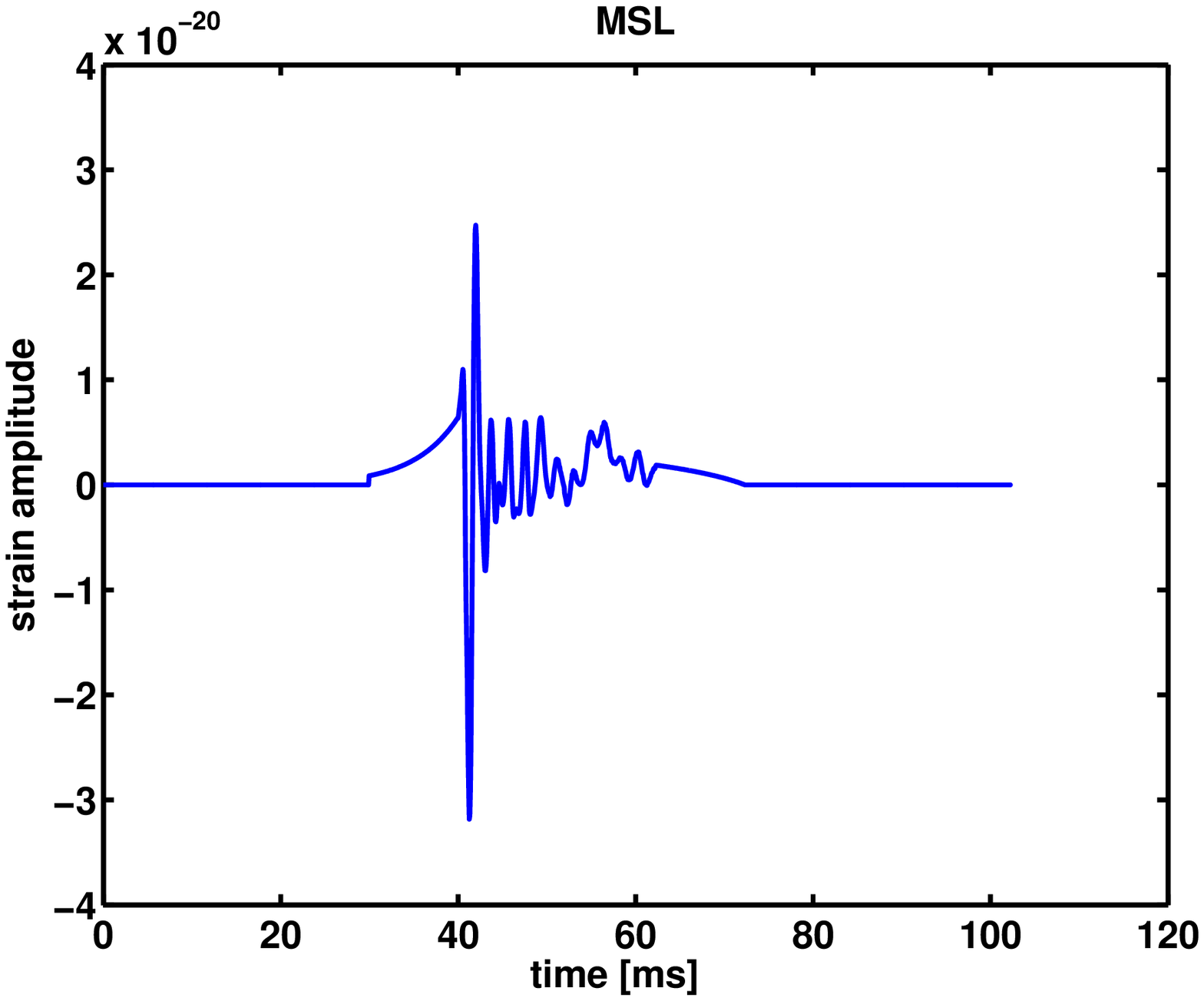}
\caption{The figures represent waveforms for various core rotational models~\cite{Kotake:2003}, as discussed in the text. Top left, top right and bottom left plots in the figure show the waveforms of the model MCS, RCS, SCS supernova signals with the optimal inclination. The bottom right plot is a signal from a model with a shell type rotation law and long differential rotation which means a small amount of differential rotation.
\label{fig:KKwaveform}}
\end{center}
\end{figure}
We take the waveforms from Kotake et al. \cite{Kotake:2003}, in which 
a series of two-dimensional Newtonian hydrodynamic simulations of the 
rotational collapse of a supernova core was done using a realistic equation of state and a treatment of  neutrino transport via the leakage 
scheme. 
Since we know little 
about the angular momentum 
distributions in the core of evolved massive stars, their simulations assumed two possible rotation laws, 
a shell-type and a cylindrical rotation law. Twelve models of gravitational waves were computed by changing the combination of total angular momentum: S (slow), M (moderate), R (rapid), the rotation law: S (shell-type), C (cylindrical), and the degree of differential rotation: L (long), S (short), which indicates the distance from the rotational axis for the cylindrical rotation or the distance from the radius for the shell-type rotation. From the simulations, they found that the sign of the peak with the second largest absolute amplitude is negative for models with strong differential rotation and with a cylindrical rotational profile in the core (model MCS, RCS, SCS) and positive for the others. Figure~\ref{fig:KKwaveform} shows the waveforms for model MCS (top left plot), RCS (top right plot), SCS (bottom left plot), and MSL (bottom right plot).
This observation implies that we can obtain information about the angular momentum distribution 
of massive evolved stars if we can detect the signs of the second peaks. This feature of the second peak also appears in simulations by other groups
working on simulations of core-collapse supernovae (See figure~6 in~\cite{ZwergerMueller:1997}, figure~11 in~\cite{Ottetal:2004}, figure~10 in~\cite{ShibataSekiguchi:2004}, figure~8 in~\cite{CerdaDuranetal:2005}. {\footnote {However, according to \cite{dimmellast}, the type II waveforms tend to transform to the type I when the general relativity and deleptonization effects are taken into account.}})

We briefly describe how the sign of the peaks is determined. 
Generally, the sign of a gravitational waveform becomes negative right before 
 core bounce because the oblateness of the core increases rapidly. 
After the bounce, the sign becomes positive as the oblateness decreases due to 
 the shock propagating from the inner core surface outwards 
along the rotational axis. 
This explains the first largest peak amplitude and origin of the negative 
sign of the first peak. After the first peak, the core oscillates with decaying 
absolute amplitude because of its inertia, known as a ring-down phase. 
In the ring-down phase, the absolute amplitudes of the subsequent peaks are 
much smaller than the first peak described above. 
After the first peak, the core expands and then begins to shrink again. 
The core's change from expansion to contraction 
 causes the positive peak in the waveform, which we call {\it{peak 1}}. 
Afterwards the amplitude begins to decrease
  until the core rebounds again. The rebound causes the subsequent 
negative peak in the waveform,  which we call {\it{peak 2}}. 
The sign of the second largest peak is determined by which absolute amplitude 
({\it{peak 1}} or {\it{peak 2}}) is larger. This is sensitive to the degree of 
 differential rotation of the core. For stronger differential 
rotation,  the centrifugal force becomes so strong that  
the gravitational contraction of the core becomes weak.  This leads to a 
weaker core-bounce than for smaller differential rotation, given the same initial 
rotation rate.
Thus {\it{peak 1}} becomes smaller as the differential rotation becomes stronger. 
Moreover, the influence of differential rotation 
can become more prominent for models with an initial cylindrical rotation profile,  which is expected to be a natural angular momentum distribution inside the core \cite{tass:1978}.
These factors create a tendency for the absolute amplitude of the 
negative peak ({\it{peak 2}}) to be larger than that of the positive peak 
({\it{peak 1}}) for stronger differential rotation with the cylindrical rotation profile.

\section{{\tt RIDGE} coherent network analysis}
 {\tt RIDGE} is a coherent network analysis pipeline described  in ~\cite{Hayama:2007}. The pipeline consists of two main components, which are data conditioning and generation of detection statistics. 

The aim of the data conditioning is to whiten the data, so as to remove frequency dependence and any instrumental artifacts in the data. 
In {\tt RIDGE}, the data are whitened by estimating the noise floor using a running median~\cite{Mukherjee:2003}. The whitening step above is followed by a line estimation and removal step by using a method described in ~\cite{Mohanty:2002}. The resulting conditioned data are passed on to the next step consisting of the generation of a detection statistic. The basic algorithm implemented for the coherent network analysis in {\tt RIDGE} is Tikhonov regularized
 maximum likelihood, described in ~\cite{Rakhmanov:2006}. Recent studies~\cite{Klimenko+etal:2005,Rakhmanov:2006,Mohanty+etal:2006} show that the inverse problem of a response matrix of a detector network becomes an ill-posed one and the resulting variance of the solution becomes large. This comes from the rank deficiency of the detector response matrix. The strength of the rank deficiency depends on the sky location, and therefore, plus or cross-polarized gravitational wave signals from some directions on the sky become too noisy.  In {\tt RIDGE}, we reduce this ill-posed problem by posing the Tikhonov regulator, which is a function of the sky location.  
The input to the algorithm is a set of equal length, conditioned data segments from the detectors in a given network. The output, for a given sky location $\theta$ and $\phi$, is the value of the likelihood of the data maximized over all possible $h_+$ and $h_\times$ waveforms with durations less than or equal to the data segments.
The maximum  likelihood values are obtained as a function of $\theta$ and $\phi$ -- this two dimensional output, ${\bf S}(\theta,\phi)$,
 is called a {\em skymap}. We can construct a detection statistic called the {\it radial distance} which uses the 
entire skymap and scales the maximum likelihood by the mean location of the same quantities in the absence of a signal.  A detailed description of the detection statistic is given in ~\cite{Hayama:2007}.

\section{Simulations and results}
\begin{figure}
\begin{center}
\includegraphics[scale=0.4]{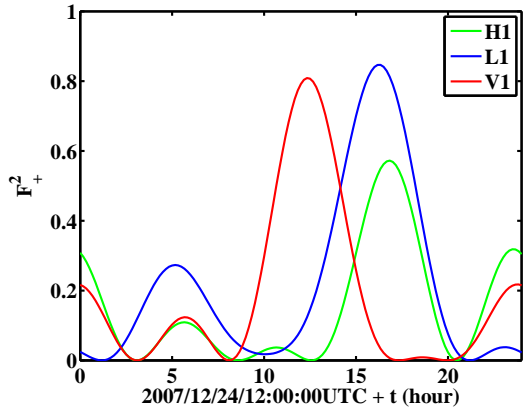}
\caption{The detector antenna function of H1, L1, V1 for the Galactic center around 3:00 am, December 25th in 2007. Here we take as the Galactic center,  the sky location at RA = 16.3 hr, DEC = $-15.8^{\circ}$. The root mean squared detection response of $h_+$ is 0.57. 
\label{fig:detrespGC}}
\end{center}
\end{figure}

\begin{figure}
\begin{center}
\includegraphics[scale=0.5]{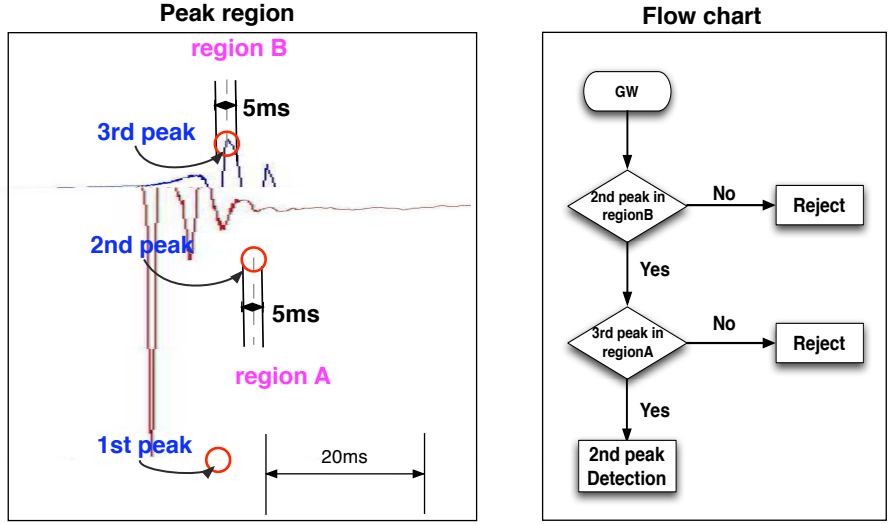}
\caption{Left figure shows the peaks and their regions. Right plot shows the flow chart for the detection of the second peak. First, we find the second peak and the third peak in a gravitational wave. When the second peak of the reconstructed waveform is within 5 ms of the true second peak, the algorithm checks whether the third peak is within 5 ms of the true third peak. If this condition is satisfied, we declare that the detection of the sign of the second peak is successful.
\label{fig:SPFindmethod}}
\end{center}
\end{figure}

We carried out Monte Carlo simulations to study the detectability of the second peak of a supernova signal using {\tt RIDGE}. Gravitational waves used here are ``+'' mode polarization waves and the models used were the MCS, RCS, SCS 
from~\cite{Kotake:2003}. We assume the rotational axis of the core is perpendicular to the direction to the Earth from the supernova so that gravitational waves are most strongly emitted towards Earth.
 We assume the rotational axis of the core is perpendicular to the direction to the earth from the supernova so that the gravitational waves are most strongly emitted. The location of the source was assumed to be that of the galactic center. The network consisted of the three LIGO detectors (Hanford/Livingston) and the VIRGO detector. All data streams are white Gaussian noise with the same sensitivities so that we focus on the performance of the signal reconstruction of the coherent network analysis. Therefore we need to add data conditioning for realistic analysis.
We assumed that the detection of the supernova signal occurred 3:00 am, December 25th in 2007, i.e. December 25th 2007 03:00. This particular event time was chosen so both the VIRGO and LIGO detectors have reasonable sensitivities to this signal. Figure~\ref{fig:detrespGC} shows the detector response for the supernova signal. The X-axis represents time and the Y-axis shows the sum-squared detector response. Simulated data streams are generated at a sampling frequency of 16384~Hz. First, the pipeline down-sampled all the data streams by applying anti-aliasing filters, resulting in a sampling frequency of 2048~Hz. In regard to the time-frequency granularity, the single tiles are $(\delta f, \delta t)=(1024~{\rm Hz}, 0.49~{\rm ms})$. The supernova signal was injected at every other second, and, in all, 200 signals were injected. The data was passed on to the {\tt RIDGE} algorithm. In {\tt RIDGE}, the data conditioning step was omitted since the noise was white Gaussian. The data was passed on to the step of the regularized coherent network analysis, and then, the estimated polarization waveforms $h_+$ were reconstructed. Picking up the reconstructed data around the injected signals we find the second largest peak of absolute amplitude. 
If the positions of the second peak and the third peak are within $5$ ms of the center of the true peak positions and the sign is negative, we argue that the detection of the sign of the second peak succeeds, otherwise the detection fails. This efficiency is characterized by a detection probability which is defined as the ratio of the number of the detections to the total number of the injected signals. This procedure is repeated by changing signal-to-noise ratio (SNR) of the supernova signals, which is the optimal output of the matched filter method. In this algorithm, we did not include the step to find the peak positions because SNR of signals is large enough to distinguish the difference of the second peak and the third peak and the positions of the peak are seen clearly.

\begin{figure}
\begin{center}
\includegraphics[scale=0.35]{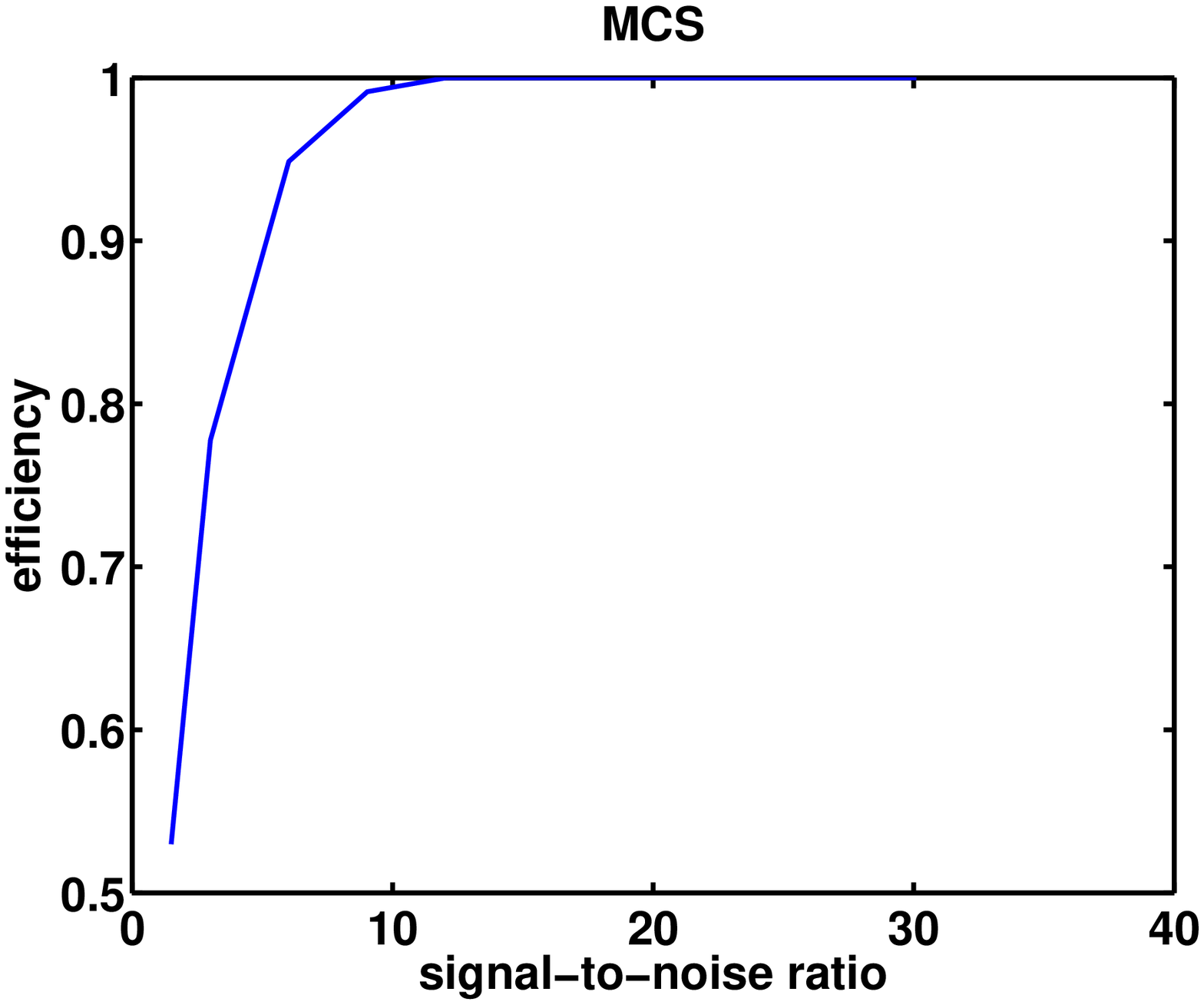}
\includegraphics[scale=0.35]{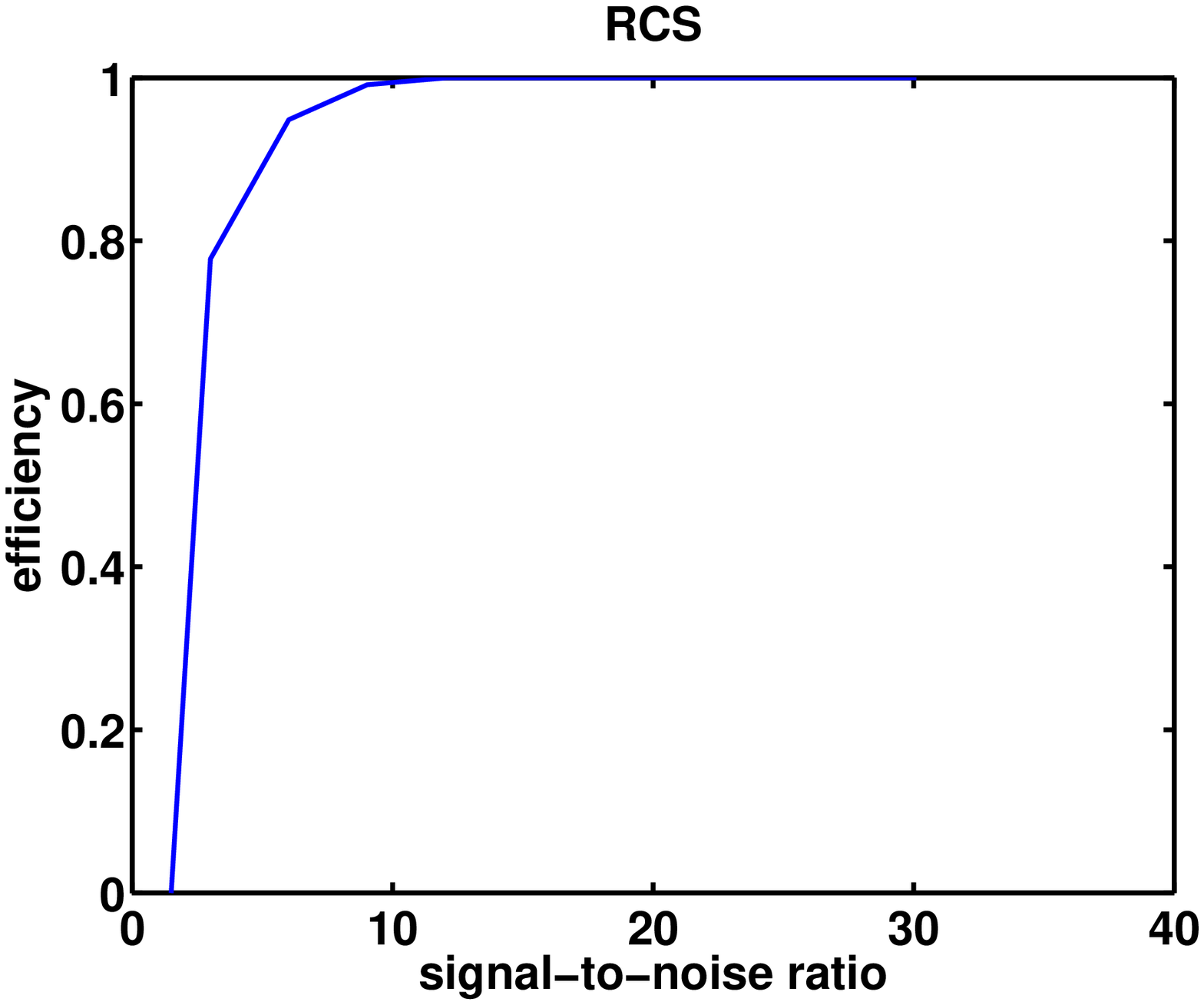}
\includegraphics[scale=0.35]{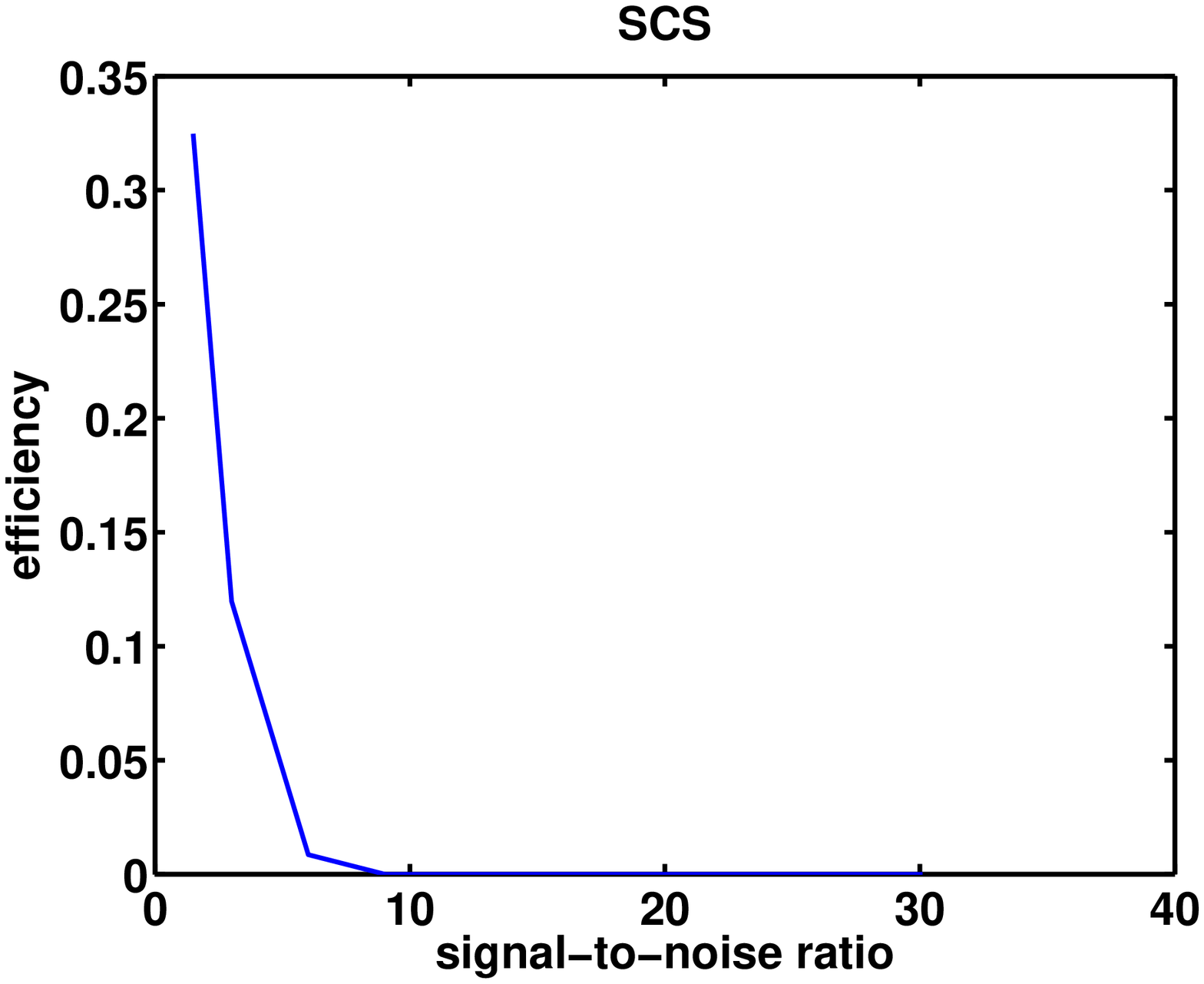}
\caption{Detection probabilities of the second peak of the model MCS supernova at the galactic center is the left upper plot, the model RCS is the right upper plot, and the model SCS is the bottom plot in the figure. The detection probability is defined as a ratio of the number of the ``-'' sign detection divided by the total number of trials which is equal to 200. The signal-to-noise ratio on the X-axis is averaged over all detectors.
\label{fig:3hChristmas}}
\end{center}
\end{figure}

Figure~\ref{fig:3hChristmas} shows the detection probability of the second peak as a function of the averaged signal-to-noise ratio over all detectors. From this figure, at 3:00 am, December 25th in 2007, the second peak of the signal MCS, RCS with signal-to-noise ratio $\ge 6$ is detectable with $\ge 90$\% efficiency. The signal-to-noise ratio of the 
supernova signals at the location of Galactic center for  model MCS is : 6.8 for H1, 5.5 for H1, 13.4 for L1, 9.6 for V1, and for model RCS: 8.9 for H1, 6.4 for H2, 12.7 for L1, 4.9 for V1. This shows that the detection of the supernova signals at the Galactic center enables us to infer the dynamics of the supernova core, particularly the rotation law and the degree of the differential rotation. Here, signal-to-noise ratio is the output obtained from matched filtering and is calculated by using design sensitivity of each detector~\cite{LIGOdesignsens,VIRGOdesignsens}. 

\begin{figure}
\begin{center}
\includegraphics[scale=0.5]{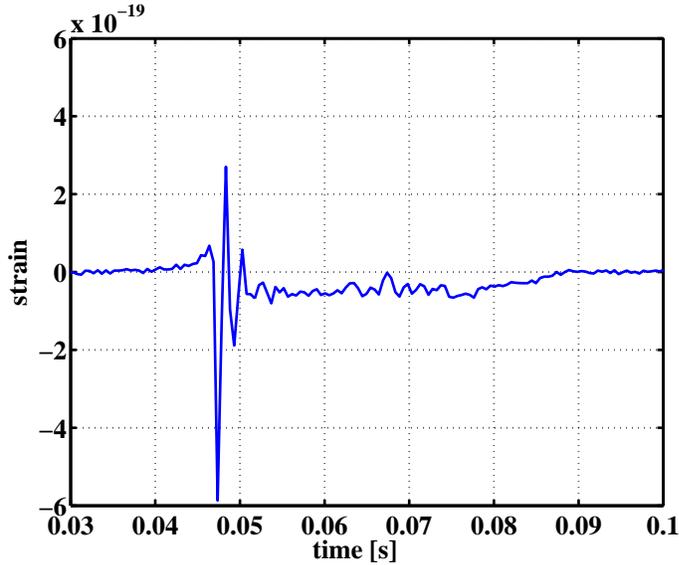}
\caption{The reconstructed waveform of model SCS. The signal-noise-ratio is 30.
\label{fig:SCSrecon}}
\end{center}
\end{figure}

Regarding the signal SCS, the detection probability is reduced as signal-to-noise ratio increases in spite of the the fact that the signal-to-noise ratio of model SCS 
is 3.8 for H1, 3.0 for H2, 8.2 for L1, 6.8 for V1. When signal-to-noise ratio is small, accidental detections occur. This is the reason why detections occur in the region with low signal-to-noise ratio. Since accidental detections become smaller and true detections become larger as signal-to-noise ratio increases, the detection efficiency becomes smaller. This result shows the second peak of the signal SCS cannot detected by the LIGO-VIRGO network at this sky location (See figure~\ref{fig:SCSrecon}).

\begin{figure}
\begin{center}
\includegraphics[scale=0.4]{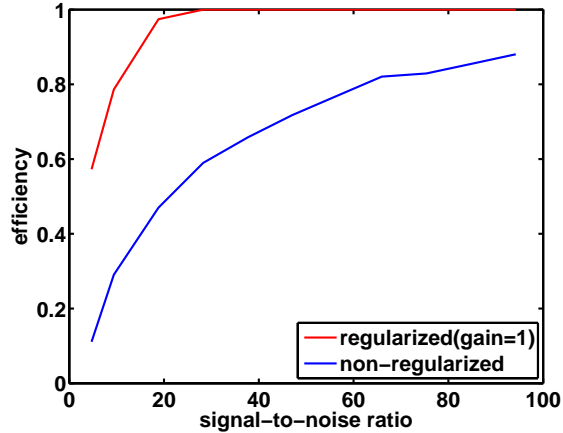}
\caption{Detection probability of second peak of the type MCS supernova signals at SN1987A as a function of a signal-to-noise ratio. The 'gain' in the plot is a scalar value which controls the strength of the regulator, taking 
 values in the range $[0,1]$. The efficiency obtained by the regularized coherent network analysis is significantly superior to that of the non-regularized one.
\label{fig:effreg}}
\end{center}
\end{figure}

Figure ~\ref{fig:effreg} shows the effect of the regulator on the detection of the second peak. A supernova signal with the model MCS was simulated assuming it arrives from the direction of SN 1987A. Assuming the event occurs at 0:00 am, December 25th in 2007, and other conditions being the same as the previous simulations, we calculated the detection efficiency. The red line in the figure is the result obtained by the regularized coherent network 
analysis with regulator gain equal to 1. The regulator gain is a scalar value taking on values in the  range $[0,1]$ and controls the strength of the regulator. The blue line is the result obtained by the non-regularized coherent network analysis, i.e., the regulator gain being zero. Since the regulator works to reduce variance of the extracted signals, it plays a significant role in the astrophysical interpretation of detected signals.

\section{Conclusion}
From recent simulations of the dynamics of a supernova core, one finds that for a cylindrical rotation law with strong differential rotation, the sign of the peak with the second largest amplitude is negative. In this paper, we performed Monte Carlo simulations to study the detection efficiency  of the sign of 
the second peak using the supernova models MCS, RCS, SCS, and
using a coherent network analysis method called {\tt RIDGE}.
Using an idealized network of GW detectors, the second peak of the
signal MCS, RCS with signal-to-noise ratio $\ge 6$ is detectable with $\ge 90$\%
 for the specific source sky location we considered. These results
are encouraging and indicate that we may be able to obtain information
about the dynamics of the supernova core, if supernova events occur
within our Galaxy. The final answer requires refined simulations.
We are in progress for this research including a part of
the data conditioning. 
We also showed that the regulator imposed on the standard coherent network analysis plays an important role in improving the detection efficiency of the second peak.

In figure \ref{fig:KKwaveform}, a low frequency component, which is like a direct current (DC) component, can be observed in the waveforms. This component appears in ring-down phase. In a more realistic case, this  component will affect the detection efficiency of the second peak since this component is difficult to see due to seismic noise. This component comes from gravitational waves radiated from material undergoing convection etc in the core and this component has negative sign \cite{CerdaDuranetal:2005, ZwergerMueller:1997}. So we can still say the negative second peak indicate strong differential rotation of the core compared with signals having a positive second peak. 

\ack{We would like to thank R Frey, J Beacom, K Sato and E M{\" u}ller  for discussions and fruitful comments on this 
paper. This work is supported by NASA NAG5-13396 to the Center for  Gravitational Wave Astronomy at the University of Texas 
at Brownsville, NSF-HRD0734800, NSF PHY-0555842, the Center for Gravitational Wave Physics (PSU), NSF 428-51 29NV0, the Office of Scholarly Research(Andrews), NSF PHY 0653233. This paper has been assigned LIGO Document Number  P080031 
}

\section*{References}

\bibliographystyle{iopart-num}
\bibliography{secondpeak}

\end{document}